\documentclass[final]{IEEEtran}

\usepackage{cite}

\usepackage{fancyhdr,comment,setspace,algorithm,algorithmic,color,mathdots}
\usepackage{amsmath,amssymb,amsthm,graphicx,graphics,subfigure,epsfig,enumerate}
\usepackage{epstopdf}
\theoremstyle{plain}\newtheorem{lemma}{\textbf{Lemma}}\newtheorem{theorem}{\textbf{Theorem}}

\theoremstyle{definition}

\newcommand{\Herm}{\mathsf{H}}
\newcommand{\Trans}{\mathsf{T}}

\newcommand{\cA}{{\mathcal A}}

\newcommand{\ba}{\boldsymbol{a}}
\newcommand{\bq}{\boldsymbol{q}}
\newcommand{\bs}{\boldsymbol{s}}

\newcommand{\be}{\boldsymbol{e}}
\newcommand{\bb}{\boldsymbol{b}}

\newcommand{\cF}{\mathcal{F}}
\newcommand{\cQ}{\mathcal{Q}}

\newcommand{\Tr}{\mathrm{Tr}} 
 
\newcommand{\argmin}{\mathrm{argmin}} 
 
\newcommand{\bg}{{\boldsymbol g}}

\newcommand{\bA}{{\boldsymbol A}}

\newcommand{\bS}{\boldsymbol{S}}
\newcommand{\bZ}{{\boldsymbol Z}}

\newcommand{\bx}{{\boldsymbol x}}
\newcommand{\by}{{\boldsymbol y}}

\newcommand{\bw}{{\boldsymbol w}}

\newcommand{\cT}{{\mathcal T}}

\newcommand{\bQ}{{\boldsymbol Q}}

\newcommand{\bz}{{\boldsymbol z}}
\newcommand{\bu}{{\boldsymbol u}}
\newcommand{\bv}{{\boldsymbol v}}
\newcommand{\bI}{{\boldsymbol I}}
\newcommand{\bY}{{\boldsymbol Y}}

\newcommand{\bX}{{\boldsymbol X}}

\newcommand{\bW}{{\boldsymbol W}}

\definecolor{hf}{RGB}{125,0,0}
\newcommand{\hf}[1]{\textcolor{black}{#1}}

\usepackage{bm}

\usepackage{algorithmic}

\usepackage{array}

\begin{document}

\title{Quantized Spectral Compressed Sensing: Cramer-Rao Bounds and Recovery Algorithms}

\author{Haoyu Fu,~\IEEEmembership{Student Member,~IEEE}, Yuejie Chi,~\IEEEmembership{Senior Member,~IEEE}
\thanks{H. Fu is with
Department of Electrical and Computer Engineering,
 The Ohio State University, Columbus, Ohio 43210.
Email: fu.436@osu.edu.}
\thanks{Y. Chi is with
Department of Electrical and Computer Engineering,
Carnegie Mellon University, Pittsburgh, PA 15217.
Email: yuejie.chi@cmu.edu.}
\thanks{Preliminary results have been presented in part at the 2017 Asilomar Conference on Signals, Systems, and Computers. }
\thanks{This work is supported in part by NSF under the grants
ECCS-1462191, CCF-1527456, ECCS-1818571, by ONR under the grant N00014-15-1-2387, and by AFOSR under the grant
FA9550-15-1-0205.}}

\maketitle

\begin{abstract}
Efficient estimation of wideband spectrum is of great importance for applications such as cognitive radio. Recently, sub-Nyquist sampling schemes based on compressed sensing have been proposed to greatly reduce the sampling rate. However, the important issue of quantization has not been fully addressed, particularly for high-resolution spectrum and parameter estimation. In this paper, we aim to recover spectrally-sparse signals and the corresponding parameters, such as frequency and amplitudes, from heavy quantizations of their noisy complex-valued random linear measurements, e.g. only the quadrant information. We first characterize the Cram\'er-Rao bound under Gaussian noise, which highlights the trade-off between sample complexity and bit depth under different signal-to-noise ratios for a fixed budget of bits. Next, we propose a new algorithm based on atomic norm soft thresholding for signal recovery, which is equivalent to proximal mapping of properly designed surrogate signals with respect to the atomic norm that motivates spectral sparsity. The proposed algorithm can be applied to both the single measurement vector case, as well as the multiple measurement vector case. It is shown that under the Gaussian measurement model, the spectral signals can be reconstructed accurately with high probability, as soon as the number of quantized measurements exceeds the order of $K\log n$, where $K$ is the level of spectral sparsity and $n$ is the signal dimension. Finally, numerical simulations are provided to validate the proposed approaches.    
\end{abstract}

 

\section{Introduction}

Emerging applications in wireless communications, cognitive radio, and radar systems deal with signals of wideband or ultrawideband \cite{tian2007compressed}. Spectrum sensing or signal acquisition in this regime is a fundamental challenge in signal processing, since the well-known Shannon-Nyquist sampling rate may become prohibitively high in practice. Therefore, it is highly desirable to come up with alternative approaches that have less demanding sampling requirements. 

Recently, compressed sensing (CS) \cite{donoho2006compressed,CandTao06} has emerged as an effective approach to allow sub-Nyquist sampling \cite{tropp2010beyond,mishali2011xampling,wakin2012nonuniform} when the wideband signal is approximately sparse in the spectral domain. The resulting paradigm is referred to as {\em Compressive Spectrum Sensing} \cite{zeng2011distributed,polo2009compressive}. Significant focus has been put on reducing the sampling rates of the analog-to-digital converters (ADC), 
which only covers one aspect of the operations of ADCs. Quantization, which maps the analog samples into a finite number of bits for digital processing, is another necessary step that requires careful treatments. Most existing works, with a few exceptions, assume that the samples are quantized at a high {bit level} so that the quantization error is relatively small and well-behaved. 

This paper aims at understanding the fundamental limits of quantization, as well as developing computationally efficient algorithms, for compressive spectrum sensing and parameter estimation, in particular in the regime of {\em heavy} quantization where it is no longer appropriate to model quantization errors as  bounded additive noise. Examining the figure-of-merit of ADCs, two key specifications are the sampling rate and the effective number of bits (ENOB), which is the number of bits per measurement, also known as the {\em bit depth}. Typically, a small bit depth allows a high sampling rate, and vice versa \cite{walden1999analog}. Therefore, it is critical to \hf{understand} the fundamental trade-off between sampling rate and bit depth for high-resolution spectrum estimation. Though the importance of understanding such trade-off has been realized in the context of CS \cite{laska2011democracy,slawski2016linear}, they haven't been studied for the task of {\em parameter estimation} using estimation-theoretic tools.

Another motivating application is wideband spectrum sensing in bandwidth-constrained wireless networks \cite{frugal2013,chi2017subspace}. In order to reduce the communication overhead, each sensor transmits quantized messages, e.g. 1-bit messages; and it is necessary to estimate wideband spectrum from quantized measurements at the fusion center. Moreover, the quantization scheme might be unknown, due to lack of the knowledge of noise statistics or privacy constraints. Therefore, it is necessary to develop estimators that do not require exact knowledge of the quantizers.


\subsection{Our Contributions}

We study high-resolution spectrum estimation of sparse bandlimited signals from quantizations of their noisy random linear measurements. The signals of interest are spectrally sparse, which contain a linear superposition of complex sinusoids with continuous-valued frequencies in the unit interval. In the extreme 1-bit case\footnote{Throughout this paper, we measure the bit depth as the number of bits used to quantizer a real number.}, the quantization is based on the quadrants of the complex-valued measurements. More generally, sophisticated quantization schemes such as Lloyd's quantizer \cite{lloyd1982least} can be used to allow a higher bit depth. The specific form of the quantizer can be either known or unknown. In addition, the quantized measurements may be additionally contaminated by a noise model to be described later, in order to model imperfections in the quantization.

In this paper, we first derive the Cram\'er-Rao bound (CRB) for estimating multiple frequencies and their complex amplitudes assuming additive white Gaussian noise (AWGN) and the Lloyd's quantizer, using a fixed \hf{and deterministic} CS measurement matrix. Our bounds suggest that the CRB experiences a phase transition depending on the signal-to-noise ratio (SNR) {\em before} quantization. In the low SNR regime it is {\em noise-limited}, and behaves similarly as if there was no quantization; in the high SNR regime, it is {\em quantization-limited}, and experiences severe performance degeneration due to quantization. Furthermore, we use the derived CRB to answer the following question: given the same budget of bits, should we use more measurements (high sample complexity) with low bit-depth, or fewer measurements (low sample complexity) with high bit-depth? We answer this question by comparing 1-bit versus 2-bit quantization schemes using the CRB, and demonstrate the answer depends on the SNR. At low SNR, 1-bit measurements are preferred, while at high SNR, 2-bit measurements are preferred.

It is well-known that maximum likelihood estimators approach the performance of CRB asymptotically at high SNR \cite{van2004detection}, however, their implementation requires exact knowledge of the likelihood function, which in our problem, includes the exact form of the quantizer and noise statistics. However, such knowledge may not be available in certain applications. Therefore, our goal is to develop estimators that do not require the knowledge of the quantization scheme. To mitigate basis mismatch \cite{chi2011sensitivity}, atomic norm \cite{chandrasekaran2012convex,tang2012compressed,chi2015compressive,li2016off,heckel2016generalized,yang2014exact,tang2015near,bhaskar2013atomic} has been proposed recently to promote spectral sparsity via convex optimization without discretizing the frequencies onto a finite grid, which has found applications in signal denoising, interpolation of missing data, and frequency localization of spectrally-sparse signals. Existing atomic norm minimization algorithms assume unquantized measurements that are possibly contaminated by additive noise, and a direct application will lead to highly sub-optimal performance when a significant amount of the quantized measurements {\em saturate} \cite{gray1971quantization}.

In this paper, we propose a novel atomic norm soft thresholding (AST) algorithm \cite{bhaskar2013atomic} to recover spectrally-sparse signals and estimate the frequencies from their 1-bit quantized measurements. Our algorithm is based on finding the proximal mapping of properly designed surrogate signals, that are formed by linear combinations of the sample-modulated measurement vectors, with respect to the atomic norm to promote spectral sparsity. In other words, we aim to find signals that balance between the proximity to the surrogate signals and the small atomic norm. Moreover, the frequencies can be localized without knowing the model order a priori, by examining the peak of a dual polynomial constructed from the dual solution. Alternatively, conventional subspace methods can be used to estimate the frequencies using the recovered spectral signal. The proposed algorithm can be generalized to handle quantizations of noisy random linear measurements of multiple spectrally-sparse signals \cite{li2016off}, where each signal contains the same set of frequencies with different coefficients. The proposed algorithms do not require knowledge of the specific form of the quantizer, and therefore can be applied even when the quantizer is unknown.

When the measurement vectors are composed of i.i.d. complex Gaussian entries, under a mild separation condition that the frequencies are separated by $4/n$, it is shown that the reconstruction error scales as $\sqrt{K\log n/m}/\lambda$, where $K$ is the level of spectral sparsity, $n$ is the signal dimension, $m$ is the number of measurements, and $\lambda$ is a parameter that depends on the SNR before quantization, which increases as we increase SNR but saturates at high SNR. Therefore, the reconstruction error rate allows a trade-off between the sample complexity and SNR.

\subsection{Related Work} 
Our work is closely related to 1-bit compressed sensing \cite{boufounos20081,gupta2010sample,yan2012robust,plan2013one,plan2013robust,plan2016generalized,plan2014high,jacques2013robust}, which aims to recover a sparse signal from signs of random linear measurements. In particular, Plan and Vershynin \cite{plan2013one,plan2013robust,plan2016generalized,plan2014high} generalize this idea to reconstructing signals that belong to some low-dimensional set. Very recently, \cite{oymak2016fast} studied a similar setup and proposed a new algorithm using projected gradient descent. The surrogate signals used in our algorithm can be traced back to \cite{plan2014high,slawski2016linear}. The difference lies in that instead of projecting the surrogate signals directly onto some low-dimensional set, we adopt the proximal mapping of the surrogate signals with respect to the atomic norm.  
Several algorithms have been proposed in the CS literature to deal with general quantization schemes \cite{laska2011democracy,boufounos2015quantization} and nonlinear measurement schemes \cite{oymak2016fast,plan2014high}, however the focus has been on reconstruction of sparse signals in a finite dictionary, whereas our focus is on parameter estimation and reconstructing sparse signals in a parametric dictionary containing an infinite number of atoms.
 
There are also several conflicting evidence regarding the trade-offs between bit-depth and sample complexity \cite{laska2012regime,slawski2016linear} for signal reconstruction, as they may vary for different problems when using specific algorithms. In contrast, we derive the Cram\'er-Rao bound for {\em parameter estimation} using quantized compressive random measurements, which provides an estimation-theoretic benchmark for gauging the trade-off as well as benchmarking performances. Our CRB adds to existing literature of CRB calculations for 1-bit quantized single-tone frequency estimation \cite{host2000effects} as well as for parameter estimation using compressive measurements \cite{pakrooh2013analysis}.

\subsection{Paper Organization and Notations}
The rest of this paper is organized as follows. Section~\ref{sec_problem} provides the problem formulation. Section~\ref{sec_crb} presents the Cram\'er-Rao bound for parameter estimation and discusses the trade-off between bit depths and sample complexity. Section~\ref{sec_approach} presents backgrounds on the atomic norm and the proposed algorithms with performance guarantees. Section~\ref{sec_MMV} presents the extension when quantizations of multiple measurement vectors are available. Numerical experiments on the proposed algorithms are provided in Section~\ref{sec_numerical}. We conclude in Section~\ref{sec_conclusion}.  

Throughout this paper, we use boldface letters to denote vectors and matrices, e.g. $\ba$ and $\bA$. The Hermitian transpose of $\ba$ is denoted by $\ba^{\mathsf{H}}$, the transpose of $\ba$ is denoted by $\ba^{\mathsf{T}}$, and $\| \bA\|$, $\|\bA\|_{\mathrm{F}}$, $\Tr(\bA)$ denote the spectral norm, the Frobenius norm, and the trace of the matrix $\bA$, respectively. An indicator function for an event $A$ is denoted as $\mathbb{I}_{A}$. Denote 
$\cT(\bu)\in\mathbb{C}^{n\times n}$ as the Hermitian Toeplitz matrix with $\bu$ as the first column.  
Define the inner product between two vectors $\ba,\bb$ as $\langle \ba,\bb\rangle =  \ba^{\mathsf{H}}\bb$. The cardinality of a set $\mathcal{D}$ is defined as $|\mathcal{D}|$. If $\bA$ is positive semidefinite (PSD), then $\bA \succeq 0$. $\Re(y)$ and $\Im(y)$ denote the real and imaginary part of a complex number $y$, respectively. The expectation of a random variable $a$ is written as $\mathbb{E}[a]$. Define $\odot$ as entry-wise product. Throughout this paper, we use $c,c_1,c_2,\ldots$ to denote universal constants whose values may change from line to line.

\section{Problem Formulation} \label{sec_problem}

Let $\bx^{\star}\in\mathbb{C}^{n}$ be a line spectrum signal, which is composed of a small number of spectral lines, defined as
\begin{equation}
\bx^{\star} = \sum_{k=1}^K c_k \bv(f_k),
\end{equation}
where $K$ is the number of frequencies or level of sparsity, $c_k =A_k e^{j 2\pi\phi_k}\in\mathbb{C}$ is the $k$th coefficient, $A_k>0$ is the $k$th amplitude, $\phi_k\in[0,1)$ is the $k$th normalized phase, $f_k\in[0,1)$ is the $k$th frequency, and
$$ \bv(f) = \begin{bmatrix}
1& e^{j2\pi f} & \cdots & e^{j2\pi(n-1)f} \end{bmatrix}^{\Trans}.$$

In CS, we acquire a set of random linear measurements of $\bx^{\star}$, contaminated by additive complex Gaussian noise, where each measurement is given as
\begin{equation}\label{Unquanztized}
z_i =\langle \ba_i, \bx^{\star} \rangle + \sigma \epsilon_i , \quad i=1,\ldots,m, 
\end{equation}
where $m$ is the number of measurements, $\ba_i\in\mathbb{C}^{n}$'s are the measurement vectors composed of i.i.d. standard complex Gaussian entries $\mathcal{CN}(0,1)$, $\sigma$ is the noise level, and we further have  i.i.d.  $\epsilon_i\sim\mathcal{CN}(0,1)$. In a vector notation, we write
\begin{equation}\label{measurement_before_quantization}
\bz = \bA \bx^{\star} + \sigma \boldsymbol{\epsilon},
\end{equation}
where $\bA =[\ba_1, \ba_2,\ldots,\ba_m]^{\Herm} \in\mathbb{C}^{m\times n}$ is the measurement matrix, $\boldsymbol{\epsilon}=[\epsilon_1,\epsilon_2,\ldots,\epsilon_m]^{\mathsf{T}}$, and $\bz = [z_1,z_2,\ldots,z_m]^{\mathsf{T}}$. These measurements are then quantized into a finite number of bits for the ease of digital storage and processing. Denote $\mathcal{Q}(\cdot): \mathbb{R} \mapsto \mathcal{D}$ as the quantizer that quantizes a real number into a finite alphabet $\mathcal{D}$, where the {\em bit depth} is the smallest number of bits necessary to represent $\mathcal{D}$, i.e. $\hf{b^{\star}} = \min\{b\in\mathbb{Z}^+: |\mathcal{D}|\leq 2^b \}$. The quantized measurements $\by = [y_1,y_2,\ldots,y_m]^{\mathsf{T}}$ of $\bz$ are then denoted as
\begin{equation}
y_i =  \mathcal{Q}(\Re(z_i)) + j \cQ(\Im(z_i)), \quad i=1,\ldots,m,
\end{equation} 
where we apply the same quantizer $\cQ$ to both the real part and the imaginary part of the complex-valued measurement $z_i$. With slight abuse of notation, we denote the quantized measurements as
\begin{equation} \label{quantized_measurements}
\by =\mathcal{Q}(\bz).
\end{equation}
Our goal is then to recover $\bx^{\star}$, and the set of frequencies $\boldsymbol{f}=\{f_k\}_{k=1}^K$, from the quantized measurements $\by$, possibly without a priori knowing the sparsity level $K$, and the form of the quantizer $\mathcal{Q}$. 

Several choices of the quantizer are of special interest. At the extreme, we consider only knowing the quadrature information of $z_i$, where 
\begin{equation}\label{sign_quantizer}
Q(a) = \mbox{sign}(a) , \quad a\in\mathbb{R},
\end{equation}
We refer to this quantizer as the {\em one-bit quantizer}, as only a single bit is used to quantize each real number.

More generally, we consider a quantizer $\mathcal{Q}(\cdot)$ that is fully characterized by the quantization intervals $\{[t_\ell,t_{\ell+1})\}_{\ell=1}^{|\mathcal{D}|-1}$, where $t_0=-\infty$, $t_{|\mathcal{D}|}=\infty$, $\cup_{\ell=1}^{|\mathcal{D}|} [t_\ell,t_{\ell+1}) =\mathbb{R}$, as well as the representatives of each interval $\omega_\ell \in[t_\ell, t_{\ell+1})$, where 
\begin{equation}
Q(a) = \omega_\ell , \quad \mbox{if} \quad  a\in [t_\ell, t_{\ell+1}).
\end{equation}
For example, the Lloyd's quantizer \cite{lloyd1982least} belongs to this form. The choice of the quantization scheme plays an important role in determining the performance of parameter estimation.

\section{Cramer-Rao Bounds and Trade-offs}\label{sec_crb}

In this section, we study the effects of quantization on parameter estimation by deriving the Cram\'er-Rao bound assuming the quantizer, the sparsity level, and the noise level are known. In particular, the bounds are calculated for 1-bit and general quantizations, respectively, which are then used to study the trade-off between sample complexity and bit depths for a fixed bit budget. 

To begin with, we assume the set of parameters, including the frequencies, amplitudes, and phases, given as $\bm{\kappa}=  \{f_k,A_k,\phi_k\}_{k=1}^{K}\in\mathbb{R}^{3K}$, is deterministic but unknown, \hf{the measurement matrix $\bA$ is deterministic and known}. Denote the probability mass function as $p(\by|\bm{\kappa})$, which is given as
\begin{align}\label{pdf_y}
p(\by|\bm{\kappa}) = \prod_{i=1}^m p(y_i|\bm{\kappa}) = \prod_{i=1}^{m} \left[ p(\Re(y_i) |\bm{\kappa})\cdot p(\Im(y_i) |\bm{\kappa})\right], 
\end{align}
where the second equality follows from the fact that $\epsilon_i$ is proper. Moreover, let 
\begin{align}
p(\Re(y_i) |\bm{\kappa}) &= \prod_{\omega \in \mathcal{D}} p_{\Re(y_i)}(  \omega |\bm{\kappa})^{\mathbb{I}_{\{\Re(y_i)= \omega\}}} \\
p(\Im(y_i) |\bm{\kappa}) &=  \prod_{\omega \in \mathcal{D}} p_{\Im(y_i)}( \omega |\bm{\kappa})^{\mathbb{I}_{\{\Im(y_i)= \omega\}}} 
\end{align}
be the probability mass function of $\Re(y)$ and $\Im(y)$, respectively. The Fisher Information Matrix (FIM), denoted by $\bI(\bm{\kappa})\in\mathbb{R}^{3K\times 3K}$, is given as
\begin{equation}\label{FIM_def}
\bI(\bm{\kappa}) = \mathbb{E}\left[ \left( \frac{\partial \log p(\by|\bm{\kappa})}{\partial \bm{\kappa}}\right) \left( \frac{\partial \log p(\by|\bm{\kappa})}{\partial \bm{\kappa}}\right)^T\right].
\end{equation}
Note that for any $1\leq i,j\leq m$,
\begin{align}
	&\mathbb{E}\left[ \left( \frac{\partial \log p(\Re(y_i) |\bm{\kappa})}{\partial \bm{\kappa}}\right) \left( \frac{\partial \log p(\Im(y_j) |\bm{\kappa})}{\partial \bm{\kappa}}\right)^T\right] \notag\\
	&= \mathbb{E}\left[   \frac{\partial \log p(\Re(y_i) |\bm{\kappa})}{\partial \bm{\kappa}} \right] \cdot \mathbb{E}\left[ \frac{\partial \log p(\Im(y_j) |\bm{\kappa})}{\partial \bm{\kappa}} \right]^T = \bm{0}, \notag 
\end{align}
where the first equality follows from independence of $\Re(y_i)$ and $\Im(y_j)$, and the second equality follows from the fact
\begin{align}
	&\mathbb{E}\left[ \left( \frac{\partial \log p(\Re(y_i) |\bm{\kappa})}{\partial \bm{\kappa}}\right)\right]\notag \\ 
	&\ =\mathbb{E}\left[
	\sum_{\omega \in \mathcal{D}} \frac{\mathbb{I}_{\{\Re(y_i)= \omega\}}}{p_{\Re(y_i)}( \omega |\bm{\kappa})} \frac{\partial  p_{\Re(y_i)}( \omega |\bm{\kappa})}{\partial \bm{\kappa}}
	 \right] \notag\\
	 &\ =  \sum_{\omega \in \mathcal{D}} \frac{\partial  p_{\Re(y_i)}( \omega |\bm{\kappa})}{\partial \bm{\kappa}} = \frac{\partial \left( \sum_{\omega \in \mathcal{D}}  p_{\Re(y_i)}( \omega |\bm{\kappa}) \right) }{\partial \bm{\kappa}} = \bm{0}. \notag 
\end{align}
Thus, plugging \eqref{pdf_y} into \eqref{FIM_def} all cross-terms will be zero and we have    
\begin{align}\label{def_FIM}
\bI(\bm{\kappa})   
&=\sum_{i=1}^m \left[\bI_i^R(\bm{\kappa}) + \bI_i^I(\bm{\kappa}) \right],
\end{align}
 where
 \begin{align*}
& \bI_i^R(\bm{\kappa})  =  \mathbb{E}\left[ \left( \frac{\partial \log p(\Re(y_i) |\bm{\kappa})}{\partial \bm{\kappa}}\right) \left( \frac{\partial \log p(\Re(y_i)  |\bm{\kappa})}{\partial \bm{\kappa}}\right)^T\right] \\
& =\sum_{\omega\in\mathcal{D}} \frac{1}{p_{\Re(y_i)}( \omega |\bm{\kappa})}\left( \frac{\partial p_{\Re(y_i)}(\omega |\bm{\kappa})}{\partial \bm{\kappa}}\right) \left( \frac{\partial p_{\Re(y_i)}(\omega  |\bm{\kappa})}{\partial \bm{\kappa}}\right)^T,
 \end{align*}
 and $\bI_i^I(\bm{\kappa})$ can be given similarly by replacing $\Re(y_i)$ with $\Im(y_i)$.

The CRB for estimating $\bm{\kappa}$, is then given as
$ \textrm{CRB}(\bm{\kappa}) = \bI(\bm{\kappa})^{-1} $, and
the CRB for estimating the $i$th parameter in $\bm{\kappa}$, is given as $[\bI(\bm{\kappa})^{-1}]_{i,i}$.

\subsection{CRB for 1-Bit Quantization}
Our goal is then to calculate the FIM in \eqref{def_FIM}. We will explain in details the calculations for the 1-bit case. First, since $\Re(z_i) \sim \mathcal{N}(\Re(\langle \ba_i,\bx^{\star}\rangle), \frac{1}{2}\sigma^2)$, then 
\begin{align}\label{PMF}
p_{\Re(y_i)}& ( \omega |\bm{\kappa}) = \mathbb{P}\left( \omega \cdot \Re(z_{i})>0| \bm{\kappa} \right) \nonumber \\
& = \frac{1}{2} +\omega \cdot \Phi \left(\frac{ \Re(\langle \ba_i,\bx^{\star}\rangle) }{\sigma} \right)  ,\; \omega =\pm 1,
\end{align}  
where $\Phi(u)= \frac{1}{\sqrt{\pi}} \int_{0}^{u}e^{-t^2}dt$. Therefore, by the chain rule,
\begin{align}\label{derivative_P}
&\; \frac{\partial p_{\Re(y_i)}(\omega |\bm{\kappa})}{\partial \bm{\kappa}}  =\frac{\omega}{ \sigma} \Phi'\left(\frac{ \Re(\langle \ba_i,\bx^{\star}\rangle) }{\sigma} \right) \frac{\partial \Re(\langle \ba_i,\bx^{\star}\rangle)}{\partial \bm{\kappa} } \nonumber  \\
 & = \frac{\omega}{\sqrt{\pi\sigma^2}}\exp\left(- \frac{ \Re(\langle \ba_i,\bx^{\star}\rangle)^2}{\sigma^2}\right) \frac{\partial \Re(\langle \ba_i,\bx^{\star}\rangle)}{\partial \bm{\kappa} } .
\end{align}

As a short-hand notation, denote $s_i(\bm{\kappa}) = \Re(\langle \ba_i,\bx^{\star}\rangle)$ and $r_i(\bm{\kappa}) = \Im(\langle \ba_i,\bx^{\star}\rangle)$. Plug \eqref{derivative_P} into $\bI_i^R(\bm{\kappa})$, we have
\begin{align*}
\bI_i^R(\bm{\kappa}) & =\frac{4\exp\left(- 2  s_i(\bm{\kappa})^2/\sigma^2 \right)}{\pi\sigma^2\left[1 - 4\Phi^2\left(\frac{ s_i(\bm{\kappa}) }{\sigma} \right)\right] }  \left(\frac{\partial s_i(\bm{\kappa})}{\partial \bm{\kappa} }\right)\left( \frac{\partial s_i(\bm{\kappa})}{\partial \bm{\kappa} }\right)^T,
\end{align*}
and similarly, 
\begin{align*}
\bI_i^I(\bm{\kappa}) & =\frac{4\exp\left(- 2  r_i(\bm{\kappa})^2/\sigma^2 \right)}{\pi\sigma^2\left[1 - 4 \Phi^2\left(\frac{ r_i(\bm{\kappa}) }{\sigma} \right)\right] }  \left(\frac{\partial r_i(\bm{\kappa})}{\partial \bm{\kappa} }\right)\left( \frac{\partial r_i(\bm{\kappa})}{\partial \bm{\kappa} }\right)^T.
\end{align*}

\hf{As a remark, when $\sigma=0$, the amplitude of the signal cannot be recovered from the 1-bit measurements due to scaling ambiguity, and the FIM becomes singular in this case. Therefore, our expressions for CRB is valid when $\sigma\neq 0$.}

\subsection{CRB for General Quantization}
We now explain the calculation for a general quantization scheme. For $\omega_\ell\in\mathcal{D}$, and a corresponding interval $[t_\ell,t_{\ell+1})$, we have
\begin{align}
p_{\Re(y_i)}& ( \omega_{\ell} |\bm{\kappa}) = \mathbb{P}\left(  \Re(z_{i}) \in[t_\ell,t_{\ell+1})  | \bm{\kappa} \right) \nonumber \\
& =\int_{\frac{t_\ell-s_i(\bm{\kappa})}{\sigma}}^{\frac{t_{\ell+1}-s_i(\bm{\kappa})}{\sigma}} \frac{1}{\sqrt{\pi}} e^{-t^2} dt \nonumber \\
&= \Phi\left(\frac{ t_{\ell+1} - s_i(\bm{\kappa}) }{\sigma} \right) -\Phi\left(\frac{ t_{\ell} - s_i(\bm{\kappa}) }{\sigma} \right) ,
\end{align}
then, following similar arguments, we have
\begin{align*}
&\; \frac{\partial p_{\Re(y_i)}(\omega_{\ell} |\bm{\kappa})}{\partial \bm{\kappa}} \nonumber \\
 & = \frac{1}{\sqrt{\pi\sigma^2}}\left[e^{- \frac{( t_{\ell+1}-s_i(\bm{\kappa}))^2}{\sigma^2}} - e^{- \frac{( t_{\ell}-s_i(\bm{\kappa}))^2}{\sigma^2}} \right] \frac{\partial \Re(\langle \ba_i,\bx^{\star}\rangle)}{\partial \bm{\kappa} } .
\end{align*}
Therefore, define
\begin{equation}
\Gamma_i^R(\bm{\kappa}) = \sum_{\ell=1}^{|\mathcal{D}|-1}\frac{\left[e^{- \frac{( t_{\ell+1}-s_i(\bm{\kappa}))^2}{\sigma^2}} - e^{- \frac{( t_{\ell}-s_i(\bm{\kappa}))^2}{\sigma^2}} \right]^2}{\Phi\left(\frac{ t_{\ell+1} - s_i(\bm{\kappa}) }{\sigma} \right) -\Phi\left(\frac{ t_{\ell} - s_i(\bm{\kappa}) }{\sigma} \right)},
\end{equation}
and
\begin{equation}
\Gamma_i^I(\bm{\kappa}) = \sum_{\ell=1}^{|\mathcal{D}|-1}\frac{\left[e^{- \frac{( t_{\ell+1}-r_i(\bm{\kappa}))^2}{\sigma^2}} - e^{- \frac{( t_{\ell}-r_i(\bm{\kappa}))^2}{\sigma^2}} \right]^2}{\Phi\left(\frac{ t_{\ell+1} - r_i(\bm{\kappa}) }{\sigma} \right) -\Phi\left(\frac{ t_{\ell} - r_i(\bm{\kappa}) }{\sigma} \right)},
\end{equation}
we have the following theorem for the expression of FIM in light of our derivations in the previous subsection.

\begin{theorem}\label{CRB-1bit}
The Fisher information matrix $\bI(\bm{\kappa})$ for estimating the unknown parameter $\bm{\kappa}$ is given as
	\begin{align}\label{FIM}
	\bI\left(\bm{\kappa} \right) &= \frac{1}{\pi \sigma^2} \sum_{i=1}^{m}\bigg(\Gamma_i^R(\bm{\kappa})  \frac{\partial s_{i}(\bm{\kappa}  ) }{\partial \bm{\kappa}} \left(\frac{\partial s_{i}(\bm{\kappa}  ) }{\partial \bm{\kappa}}\right)^{T}\nonumber  \\  
	&\quad \quad +\Gamma_i^I (\bm{\kappa}) \frac{\partial r_{i}(\bm{\kappa}  ) }{\partial \bm{\kappa}} \left(\frac{\partial r_{i}(\bm{\kappa}  ) }{\partial \bm{\kappa}}\right)^{T}	\bigg).
	\end{align}
\end{theorem}

It is worth mentioning the FIM depends only on the quantization intervals, {\em not} the value of representatives. In contrast, the FIM using the unquantized measurements $\bz$ is given as
\begin{align}
	\bI_{\mathrm{unquantized}}\left(\bm{\kappa} \right)  &= \frac{2}{\sigma^{2}} \sum_{i=1}^{m}   \frac{\partial s_{i}(\bm{\kappa}  ) }{\partial \bm{\kappa}} \left(\frac{\partial s_{i}(\bm{\kappa}  ) }{\partial \bm{\kappa}}\right)^{T}\nonumber  \\  
	&\quad \quad +  \frac{\partial r_{i}(\bm{\kappa}  ) }{\partial \bm{\kappa}} \left(\frac{\partial r_{i}(\bm{\kappa}  ) }{\partial \bm{\kappa}}\right)^{T}	\bigg).
\end{align}
 
  \begin{figure*}[htp]
\hspace{-0.6in}\includegraphics[width=1.19\textwidth]{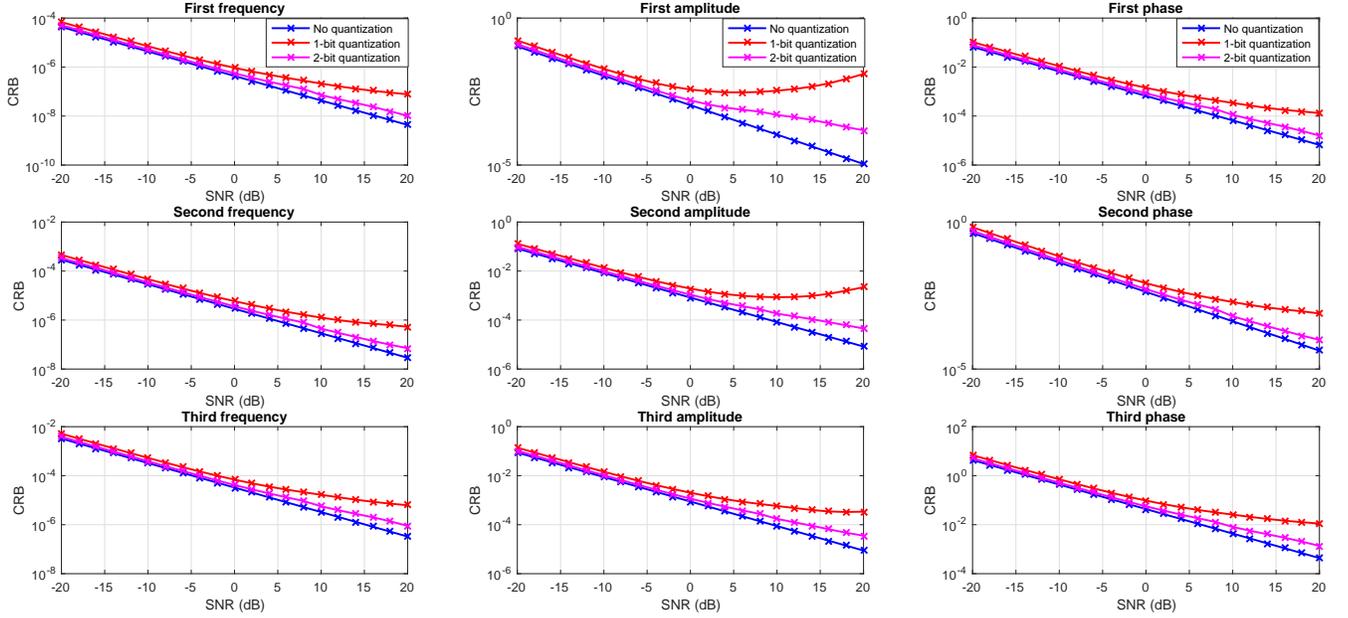}
 	\caption{The CRB under different bit-depths with respect to SNR for a fixed number of measurements $m=100$. Here, $n=64$ and $K=3$. Each row represents the CRB for estimating the frequency, amplitude and phase of one spectral atom.}\label{CRB_FixedMeasurements}
 \end{figure*}

It remains to evaluate $\frac{\partial  s_{i}(\bm{\kappa}  )}{\partial \kappa}$ and $\frac{\partial  r_{i}(\bm{\kappa}  )}{\partial \kappa}$. Following the Wirtinger calculus \cite{adali2011complex}, we have $\frac{\partial  s_{i}(\bm{\kappa}  )}{\partial \kappa}= \frac{\partial \Re(\langle \ba_i,\bx^{\star}\rangle)}{\partial \kappa} =\frac{1}{2}\Re\left(\ba_i^{\Herm}\frac{\partial \bx^{\star}}{\partial \kappa} \right)$ and $\frac{\partial  r_{i}(\bm{\kappa}  )}{\partial \kappa}= \frac{\partial \Im(\langle \ba_i,\bx^{\star}\rangle)}{\partial \kappa} =\frac{1}{2}\Im\left(\ba_i^{\Herm}\frac{\partial \bx^{\star}}{\partial \kappa} \right)$. Define 
\begin{equation*}
\boldsymbol{g}(f) = \frac{\partial \boldsymbol{v}(f)}{\partial f} = \begin{bmatrix}
0, j2\pi e^{j2\pi f} , \cdots  ,j2\pi(n-1) e^{j2\pi(n-1)f} \end{bmatrix}^{\Trans}.
\end{equation*}
Then, for each of the parameters in $\bm{\kappa}$, we have, for $k=1,\ldots,K$,
\begin{subequations} \label{derivatives_x}
\begin{align}
	\frac{\partial \bx^{\star}}{\partial f_{k}} &= c_k \boldsymbol{g}(f_k),  \\
	\frac{\partial \bx^{\star}}{\partial A_k} &= e^{j2\pi \phi_{k}}\bv(f_k),   \\
	\frac{\partial \bx^{\star}}{\phi_{k}} &=j2\pi c_k  \bv(f_k).      
\end{align}
\end{subequations}

 \begin{figure*}[htp]
\hspace{-0.6in}\includegraphics[width=1.19\textwidth]{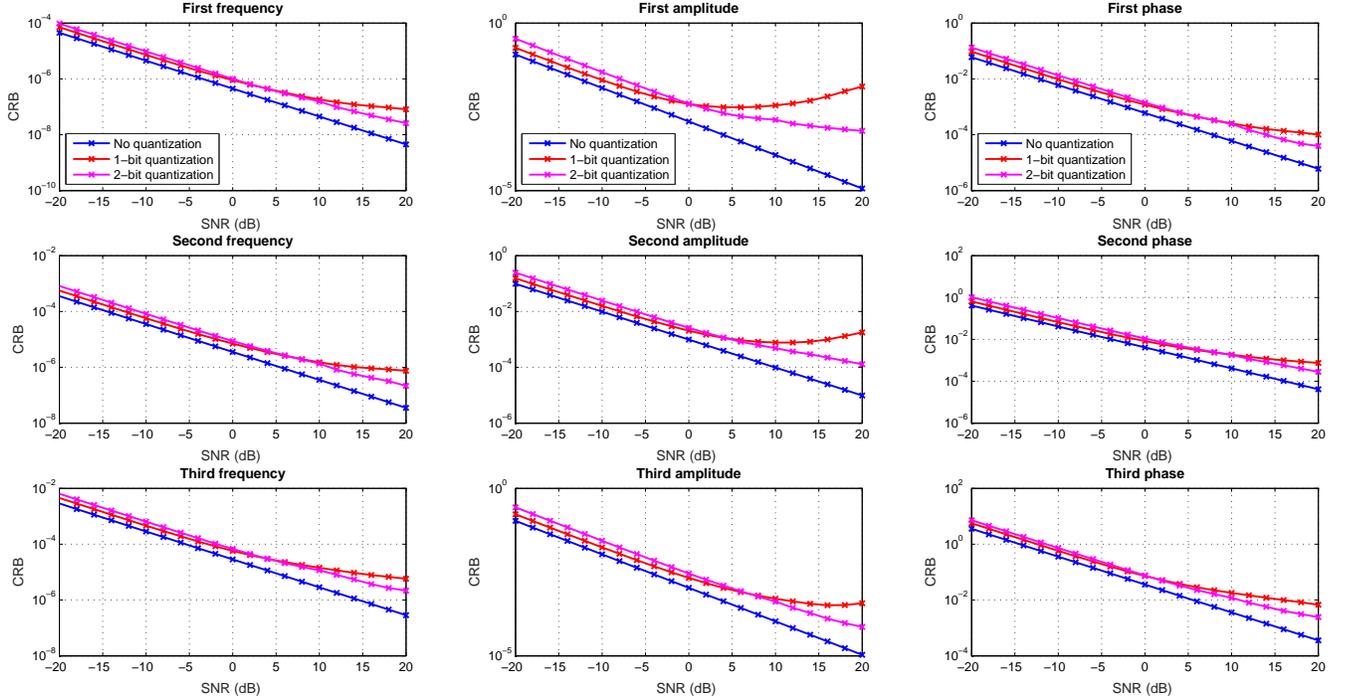} 
 	\caption{The CRB under different bit-depths with respect to SNR for a fixed number of bits $B=100$. In this case, 2-bit quantization only has half the number of measurements of the 1-bit case. Here, $n=64$ and $K=3$. Each row represents the CRB for estimating the frequency, amplitude and phase of one spectral atom.}\label{CRB_FixedBits}
 \end{figure*}
 
\subsection{Numerical Evaluations}
 
We now evaluate the CRB for 1-bit and 2-bit quantization schemes using the Lloyd's quantizer, and compare it against the CRB without quantization. We generate a spectrally-sparse signal $\bx^{\star}$ of length $n=64$ with frequencies $f_1=0.3$, $f_2=0.325$, and $f_3=0.8$, and complex coefficients $c_1=0.4e^{j2\pi\cdot 0.1}$, $c_2=0.15e^{j2\pi \cdot 0.55}$, and $c_3=0.05e^{j2\pi\cdot 0.75}$, which are selected arbitrarily.

We first fix the number of measurements as $m=100$, and generate a measurement matrix with complex standard i.i.d. Gaussian entries. Fig.~\ref{CRB_FixedMeasurements} shows the CRB for estimating all parameters with respect to the {SNR}, where it is defined as $\mathsf{SNR}= \| \bx^{\star}\|_{2}^2/\sigma^2$. It is evident that increasing the bit depth improves the performance. In the low {SNR} regime performance is {\em noise-limited}, and behaves similarly as if there was no quantization; in the high {SNR} regime, performance is {\em quantization-limited}, and experiences severe performance degeneration due to quantization. 

In many situations, we cannot simultaneously have high sample complexity and high bit depth, but rather, our budget is constrained by the number of total bits, which is $B= m\cdot \hf{b^{\star}} = m\cdot \lceil \log_2|\mathcal{D}|\rceil$. Therefore, it is useful to understand the trade-off between sample complexity and bit depth. Here, we use the CRB as a tool to compare the 1-bit and 2-bit quantization schemes. Fix the total number of bits as $B=100$. In the 1-bit quantization scheme, we use a measurement matrix with $m=100$ as generated earlier. In the 2-bit quantization scheme, we only use the first $m/2$ rows of the same measurement matrix. For comparison, we also plot the CRB assuming unquantized measurements using the same measurement matrix as the 1-bit case. Fig. \ref{CRB_FixedBits} shows the CRB for estimating all parameters with respect to the {SNR}. It can be seen that in the low {SNR} regime, 1-bit quantization is preferred, as performance is noise-limited, so higher sample complexity improves performance; in the high {SNR} regime, 2-bit quantization is preferred, as performance is quantization-limited, so higher bit depth improves performance.  Our analysis is estimation-theoretic, and doesn't depend on the algorithm being adopted.

\section{Atomic Norm Soft Thresholding for Quantized Spectral Compressed Sensing} \label{sec_approach}
It is well-known that maximum likelihood estimators approach the performance of CRB asymptotically at high SNR \cite{van2004detection}, however, their implementation requires exact knowledge of the likelihood function, which may not be available in certain applications. Therefore, in this section, we will develop estimators that do not require the knowledge of the quantization scheme using 1-bit measurements via atomic norm minimization \cite{chandrasekaran2012convex}. We first provide the backgrounds on atomic norm for line spectrum estimation, and then describe the proposed algorithms for both the single vector case and the multiple vector case with performance guarantees.

\subsection{Backgrounds on Atomic Norms}
The atomic norm is originally proposed in \cite{chandrasekaran2012convex} as a unified framework of convex regularizations for solving underdetermined linear inverse problems. Subsequently,  \cite{tang2012compressed,chi2015compressive,li2016off,heckel2016generalized,yang2014exact,tang2015near,bhaskar2013atomic} has tailored it to the estimation of spectrally-sparse signals. 

For the single vector case, define the {\em atomic set} as $\mathcal{A}_s =\left\{ e^{j\phi} \bv(f): \; f\in[0,1), \phi\in[0,2\pi) \right\}$, then the atomic norm of a vector $\bx$ is given as
 \begin{align}\label{def_atomic_norm}
  \|\bx\|_{\cA} & :=\inf\{t>0: \; \bx \in t \cdot \mbox{conv}(\cA_s) \} \nonumber \\
  & = \inf \left\{ \sum_i |\alpha_i| \; \Big| \;  \bx = \sum_i \alpha_i  \bv(f_i) \right\},
\end{align}
where $\mbox{conv}(\cA)$ denotes the convex hull of set $\cA$. The atomic norm
 can be viewed as a continuous analog of the $\ell_1$ norm over the continuous dictionary defined by the atomic set. \hf{Therefore, by promoting signals with small atomic norms, we encourage signals that can be expressed by a small number of spectral atoms.} Appealingly, as shown in \cite{tang2012compressed}, it is possible to calculate $\|\bx\|_{\cA}$ using an equivalent semidefinite program, which can be computed efficiently using off-the-shelf solvers:
$$   \|\bx\|_{\cA} = \min_{\bu\in\mathbb{C}^n, w}\left\{  \frac{1}{2n}\Tr(\cT(\bu)) +\frac{w}{2} \Big| \begin{bmatrix}
 \cT(\bu) & \bx \\
 \bx^{\Herm} & w \end{bmatrix} \succeq 0 \right\}, $$
 \hf{where $\cT(\bu)$ denotes the Hermitian Toeplitz matrix with $\bu$ as the first column.}
The dual atomic norm $\|\cdot\|_{\cA}^{*}$ for a vector $\bq\in\mathbb{C}^n$, as will become useful later, is given as
\begin{align*} 
\| \bq\|_{\cA}^* =\sup_{\|\bx\|_{\cA}\leq 1} \langle \bq, \bx \rangle_{\mathbb{R}}  &= \sup_{f\in[0,1]}   |  \bq^{\Herm}\bv(f)     |,
\end{align*}
\hf{where the second equality follows from the fact the the extreme values are taken when $\bx$ is aligned with $\bv(f)$ due to convexity}. From the above equation it is clear that $\|\bq\|_{\cA}^* $ can be interpreted as the largest absolute value of a polynomial of $e^{j2\pi f}$, denoted as $Q(f) =|  \bq^{\Herm}\bv(f)     |$.

\subsection{Atomic Soft-Thresholding with Quantized Measurements}
We first construct a surrogate signal from the quantized measurements as \cite{plan2014high}
\begin{equation}\label{surrogate}
\bs = \frac{1}{m}\sum_{i=1}^m  y_i \ba_i = \frac{1}{m}\bA^{\Herm} \by \in\mathbb{C}^n,
\end{equation}
and use the following atomic norm soft-thresholding (AST) algorithm to estimate the signal $\bx$,
\begin{equation}\label{1bit_atomic}
\hat{\bx} = \argmin_{\bx\in\mathbb{C}^n} \frac{1}{2}\|\bx - \bs \|_2^2 + \tau \|\bx\|_{\cA},
\end{equation}
which is the proximal mapping of the surrogate signal $\bs$ with respect to the atomic norm, where $\tau>0$ is a regularization parameter. One appealing feature of atomic norm minimization is that the set of frequencies can be recovered via the dual polynomial approach \cite{bhaskar2013atomic}. Namely, denote the dual variable as $\hat{\bq} = (\bs -\hat{\bx})/\tau$, and $Q(f)=| \hat{\bq}^{\Herm} \bv(f)| $. Then the set of frequencies can be localized as $\hat{\cF}=\{f: Q(f)=1\}$. We refer interested readers to the details in \cite{tang2012compressed}. Alternatively, the frequencies can be localized via performing conventional subspace methods using the estimated signal.

\subsection{Performance Guarantees}

\begin{figure}[t]
\begin{center}
\includegraphics[width=0.45\textwidth]{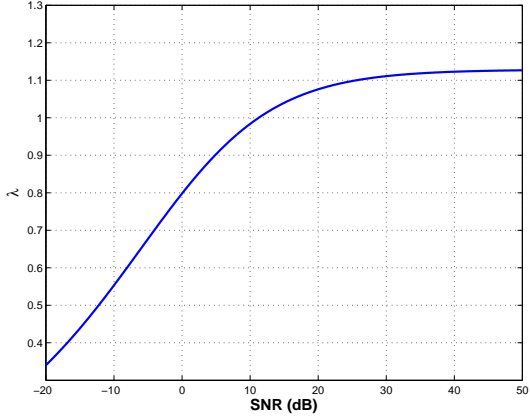} 
\end{center}
\caption{The value of $\lambda$ with respect to $\mathsf{SNR}$ before quantization.}\label{lambda_1bit}
\end{figure}

In this section, we develop performance guarantees of the proposed AST algorithm under 1-bit quantization in the single vector case using the sign quantizer in \eqref{sign_quantizer}. Note that in this case, it can be seen that $\bs$ in \eqref{surrogate} is an unbiased estimator of $\bx^{\star}$ up to a scaling difference, i.e.
\begin{equation*}
\mathbb{E}[\bs] = \lambda \frac{ \bx^{\star}}{\|\bx^{\star}\|_2},
\end{equation*}
where 
\begin{equation}
\lambda = \frac{2\|\bx^{\star}\|_2}{\sqrt{\pi(\sigma^2+\|\bx^{\star}\|_{2}^{2})}} =\frac{2}{\sqrt{\pi(1/\mathsf{SNR}+1)}}  
\end{equation}
depends on the $\mathsf{SNR}$ before quantization $\mathsf{SNR}= \|\bx^{\star}\|_2^2/\sigma^2$. To illustrate, Fig.~\ref{lambda_1bit} depicts $\lambda$ as a function of $\mathsf{SNR}$, which is a monotonically increasing function with respect to $\mathsf{SNR}$ and approaches to the limit $2/\sqrt{\pi}$ as $\mathsf{SNR}$ goes to infinity.

Without loss of generality, we assume $\|\bx^{\star}\|_2=1$. The performance of AST relies critically on the separation condition, which is defined as the minimum distance between distinct frequencies,
\begin{equation}
\Delta = \min_{k\neq j} \left\vert f_{k} - f_{j}\right\vert\ge \frac{4}{n}, 
\end{equation}
where $|f_k-f_j|$ is evaluated as the wrap-around difference on the unit modulus. Under the separation condition, we have the performance guarantee of the proposed algorithm in \eqref{1bit_atomic}, stated below.

\begin{theorem} \label{theorem_main}Set $\tau:=\eta \sqrt{n\log n/m}$ for some constant $\eta\geq 1$. Under the separation condition, the solution $\hat{\bx}$ satisfies
$$ \left\|  \frac{\hat{\bx}}{\lambda} - \bx^{\star} \right\|_2  \lesssim \frac{1}{\lambda}\sqrt{\frac{K\log n}{m}} $$
with high probability.
\end{theorem}

The proof of Theorem~\ref{theorem_main} can be found in Appendix~\ref{proof_theorem_smv}. Theorem~\ref{theorem_main} suggests that the proposed algorithm accurately recovers the signal as soon as $m$ is on the order of $K\log n$, which is order-wise near-optimal, since at least an order of $K\log (n/K)$ measurements are needed in order to recover a sparse signal in the DFT basis \cite{plan2013robust}. Moreover, the theorem also suggests that the normalized reconstruction error is inverse proportional to $\lambda$, which plays the role of SNR {\em after quantization} and is a nonlinear function of the SNR before quantization. In the low SNR regime, $\lambda$ scales as $1/\sqrt{\mathsf{SNR}}$, and the performance is comparable to that using unquantized measurements. However, in the high SNR regime, there is a saturation phenomenon, as evidenced by Fig.~\ref{lambda_1bit}, and the performance does not improve as much with we increase SNR, which is also corroborated by numerical simulation in Section~\ref{sec_numerical}. These results are qualitatively in line with existing work on one-bit CS \cite{plan2013robust}.  

\noindent\textbf{Remark:} More generally, Theorem~\ref{theorem_main} can be extended to the generalized linear model following similar strategies in \cite{plan2013one}, as long as the 1-bit measurements $y_i$'s are i.i.d. and satisfy 
$\mathbb{E}[y_i|\ba_{i}] = g\left(\langle \ba_i, \bx^{\star}\rangle \right)$
for some link function $g(\cdot)$, and accordingly $\lambda = \mathbb{E}[g(\theta)\theta^{\Herm}]$ where the expectation is taken with respect to $\theta\sim\mathcal{CN}(0,1)$. This allows us to model other complex quantization schemes with non-Gaussian noise.

\section{Extension to the Multiple Vector Case } \label{sec_MMV}

In many applications, we encounter an ensemble of line spectrum signals, where each signal $\bx_t\in\mathbb{C}^n$ contains a linear combination of spectral lines with the same set of frequencies $\mathcal{F}$, but with varying amplitudes, given as
\begin{equation*}
\bx^{\star}_{t} = \sum_{k=1}^{K}c_{k,t}\bv\left(f_{k} \right), \quad 1\leq t\leq T ,
\end{equation*}
where $c_{k,t}\in\mathbb{C}$, and $T$ is the number of snapshots. Denote $\bX^{\star}=[\bx_1^{\star},\bx_2^{\star},\ldots,\bx_T^{\star}] \in \mathbb{C}^{n\times T}$ as the signal ensemble. Similar to \eqref{measurement_before_quantization}, the CS measurement of each snapshot is given as
\begin{equation}
\bz_t = \bA \bx_t^{\star} + \sigma \boldsymbol{\epsilon}_t,
\end{equation}
where $ \boldsymbol{\epsilon}_t =[\epsilon_{1,t},\epsilon_{2,t},\ldots,\epsilon_{m,t}]^{\mathsf{T}}$ contains i.i.d. standard complex Gaussian $\mathcal{CN}(0,1)$ entries. Similar to \eqref{quantized_measurements}, the quantized measurements of each $\bz_t$ is then given as 
\begin{equation}
\by_t = \mathcal{Q}(\bz_t).
\end{equation}
Denote $\bZ=[\bz_1,\bz_2,\ldots,\bz_T]$ and $\bY=[\by_1,\by_2,\ldots,\by_T]$ as the unquantized measurement ensemble and the quantized measurement ensemble, respectively. Our goal is then to recover $\bX^{\star}$ and the set of frequencies from $\bY$, without assuming the knowledge of the sparsity level and the quantizer. The presence of multiple vectors can significantly improve the accuracy of frequency estimation.

It is possible to extend the atomic norm formulation to the multiple vector case \cite{li2016off}. Define the atomic set as 
$$\mathcal{A}_m = \left\lbrace \bA\left(f,\bb \right)=\bv\left(f \right)\bb | f\in \left( 0,1\right],\bb\in\mathbb{C}^{1\times T} ,  \|\bb\|=1 \right\rbrace, $$ 
then the atomic norm is defined as
\begin{align*}
	\|\bX\|_{\mathcal{A}} &= \mathrm{inf} \left\lbrace t>0: \bX\in t\cdot \mathrm{conv}\left( \mathcal{A}_m \right) \right\rbrace \\
	 & = \mathrm{inf} \left\lbrace \sum_{k} |c_k| \; \Big|\bX = \sum_{k} c_{k} \bA\left(f_{k},\bb_{k} \right)  \right\rbrace,
\end{align*}
which can be computed similarly via solving the following semidefinite program \cite{li2016off}:
\begin{align*} 
\|\bX\|_{\cA} = \min_{\bu\in\mathbb{C}^n, \bW\in\mathbb{C}^{T\times T}}\Big\{  \frac{1}{2n}\Tr(\cT(\bu)) +\frac{1}{2}\Tr(\bW) \Big| \\
\begin{bmatrix}
 \cT(\bu) & \bX \\
 \bX^{\Herm} & \bW \end{bmatrix} \succeq 0 \Big\}. 
\end{align*}
The dual norm for some $\bQ\in\mathbb{C}^{n\times T}$ is given as
\begin{equation*}
\|\bQ\|_{\cA}^* = \sup_{\|\bX\|_{\cA}\leq 1} \langle \bQ, \bX \rangle_{\mathbb{R}}  = \sup_{f\in[0,1]}   \|  \bQ^{\Herm}   \bv(f)  \|,
\end{equation*}
which is the largest absolute value of the polynomial $Q(f)= \|  \bQ^{\Herm}   \bv(f)  \|$.

For reconstruction, we construct the surrogate signal ensemble from the quantized measurement ensemble $\bY$ as
\begin{equation}\label{surrogate_mmv}
\bS = \frac{1}{m}\bA^{\Herm}\bY \in\mathbb{C}^{n\times T},
\end{equation}
and use the following atomic norm soft-thresholding (AST) algorithm to estimate the signal ensemble $\bX$,
\begin{equation}\label{atomic_denoising_MMV}
\hat{\bX} = \argmin_{\bX\in\mathbb{C}^{n\times T}} \| \bX - \bS \|_{F}^{2}+\tau_T \|\bX\|_{\mathcal{A}},
\end{equation}
where $\tau_T>0$ is a regularization parameter. Moreover, define $\hat{\bQ} = (\bS -\hat{\bX})/\tau_T$, and $Q(f)=\| \hat{\bQ}^{\Herm} \bv(f) \|$. Then the set of frequencies can be localized as $\hat{\cF}=\{f: Q(f)=1\}$. Alternatively, the frequencies can be localized via performing conventional subspace methods using the estimated snapshots.

\section{Numerical Experiments}\label{sec_numerical}
In this section, we conduct numerical experiments to evaluate the performance of the proposed AST algorithms for parameter estimation using quantized compressive measurements in both the single vector case and the multiple vector case. \hf{For implementation of the AST algorithms, we used the CVX toolbox \cite{grant2008cvx}. There're several other fast solvers developed for atomic norm minimization that are more scalable to large problems, including ADMM \cite{bhaskar2013atomic,li2016off}, ADCG \cite{boyd2017alternating}, and CoGent \cite{rao2015forward}, to name a few.}

\subsection{Single Vector Case} 
\begin{figure}[h]
\begin{center}
\includegraphics[width=0.45\textwidth]{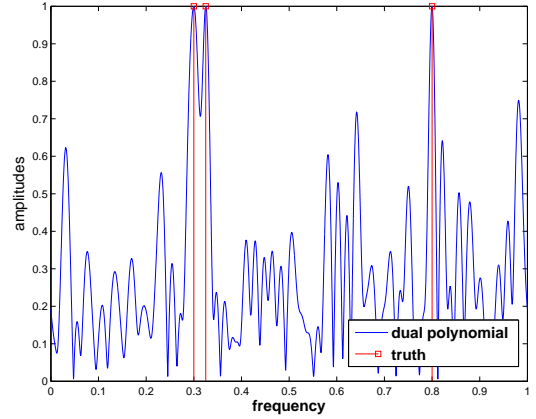} 
\end{center}
\caption{Frequency localization via peaks of the dual polynomial, superimposed on the ground truth.}\label{1bit_localization}
\end{figure}

Let $n=64$ and $K=3$. The set of frequencies is located at $\boldsymbol{f}= \{ 0.3, 0.325, 0.8\}$, where the first two frequencies are separated barely more than $1/n$, the Rayleigh limit. The number of bits is set as $m=1000$, where the measurement vectors are generated with i.i.d. $\mathcal{CN}(0,1)$ entries. The measurements are quantized according to \eqref{sign_quantizer}. Fig.~\ref{1bit_localization} shows the amplitude of the constructed dual polynomial by solving \eqref{1bit_atomic}, where its peaks can be used to localize the frequencies. It can be seen that it matches accurately with the ground truth.  

\begin{figure}[h]
\begin{center}
\includegraphics[width=0.45\textwidth]{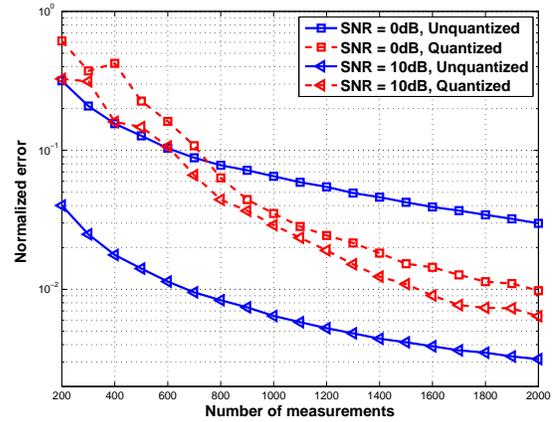} 
\end{center}
\caption{Normalized reconstruction error with respect to the number of measurements at different SNRs with or without quantization.}\label{1bit_quantization}
\end{figure}
Next, we compare the performance of signal reconstruction using atomic norm with unquantized measurements $\bz$, by running the algorithm:
\begin{equation*}
\hat{\bx}_{\mathsf{UQ}} = \argmin_{\bx\in\mathbb{C}^n} \frac{1}{2}\|\bz -\bA \bx \|_2^2 + \tilde{\tau} \|\bx\|_{\cA},
\end{equation*}
where $\tilde{\tau}$ is a properly tuned regularization parameter. The normalized reconstruction error is defined as $\sin^2(\angle \hat{\bx},\bx) = 1-|\langle \hat{\bx},\bx^{\star}\rangle|^2/(\|\hat{\bx}\|_2^2 \|\bx^{\star}\|_2^2)$, where $\hat{\bx}$ is the reconstructed signal using either algorithm. Fig.~\ref{1bit_quantization} shows the normalized reconstruction error at different SNRs with comparisons to that using the quantized measurements and the AST algorithm \eqref{1bit_atomic}, where SNR is defined again as $\mathsf{SNR}= \| \bx^{\star}\|_{2}^2/\sigma^2$. It can be seen that the reconstruction accuracy improves as we increase the SNR as well as the number of measurements, validating the theoretical analysis. In particular, at low SNR, using quantized measurements can potentially achieve better reconstruction quality with much fewer measurement budgets in bits. It can also be seen that improving the SNR before quantization does not have as strong impact as for the unquantized case.

\begin{figure}[ht]
\begin{center}
\includegraphics[width=0.45\textwidth]{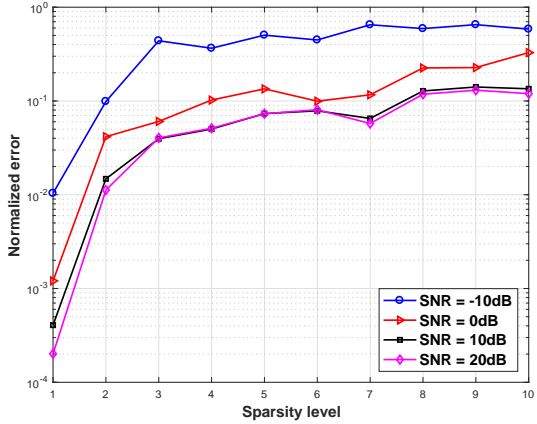} 
\end{center}
	\caption{Normalized reconstruction error with respect to the spectral sparsity level at different SNRs before quantization.}\label{1bit_SNR_K}
\end{figure}

Next, we examine the performance of the proposed algorithm as a function of the spectral sparsity level. Fix $n=64$ and $m=1000$. At each run, we randomly generate $K$ different frequencies that satisfy the separate condition. Fig.~\ref{1bit_SNR_K} shows the normalized reconstruction error as a function of the sparsity level at various SNR, averaged over $200$ Monte Carlo simulations. It can be seen that the reconstruction error is higher when the spectral sparsity level is higher, and the SNR is lower. Moreover, it can be seen that the reconstruction error stops to decrease when the SNR is relatively high, indicating a saturation effect due to quantization, as predicted by our theory.

\begin{figure}[ht]
\begin{center}
\includegraphics[width=0.45\textwidth]{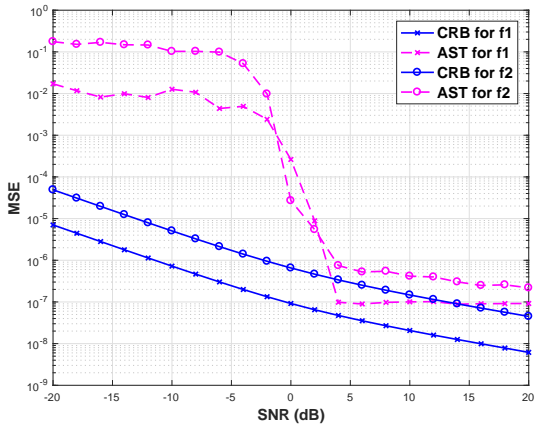}  
\end{center}
	\caption{Mean square error of frequency localization with respect to SNR using 1-bit measurements, CRB is provided as a benchmark: (a) first frequency; (b) second frequency.}\label{1bit_Performance_CRB}
\end{figure}  

We further compare the performance of frequency localization using the proposed algorithm with the CRB. Fix $n = 64$ and $m=1000$. We generate the ground signal with frequencies $f_1=0.3$, $f_2 =0.325$ and amplitudes $c_1=0.4e^{j2\pi\cdot 0.1}$, $c_2=0.15e^{j2\pi \cdot 0.55}$. Fig.~\ref{1bit_Performance_CRB} shows the average mean squared error for each frequency over 200 Monte Carlo simulations, against the corresponding CRB calculated using the formulas in Section~\ref{sec_crb}. The frequencies are estimated by using the MATLAB function \texttt{rootmusic} by assuming the correct model order, that is $K=2$. The performance of the proposed algorithm exhibits a threshold effect where it approaches that of CRB as soon as SNR is large enough. However, further increasing the SNR doesn't seem to improve the performance, which coincides with the saturation effect discussed earlier.

\begin{figure*}[ht]
\begin{tabular}{cc}
\includegraphics[width=0.45\textwidth]{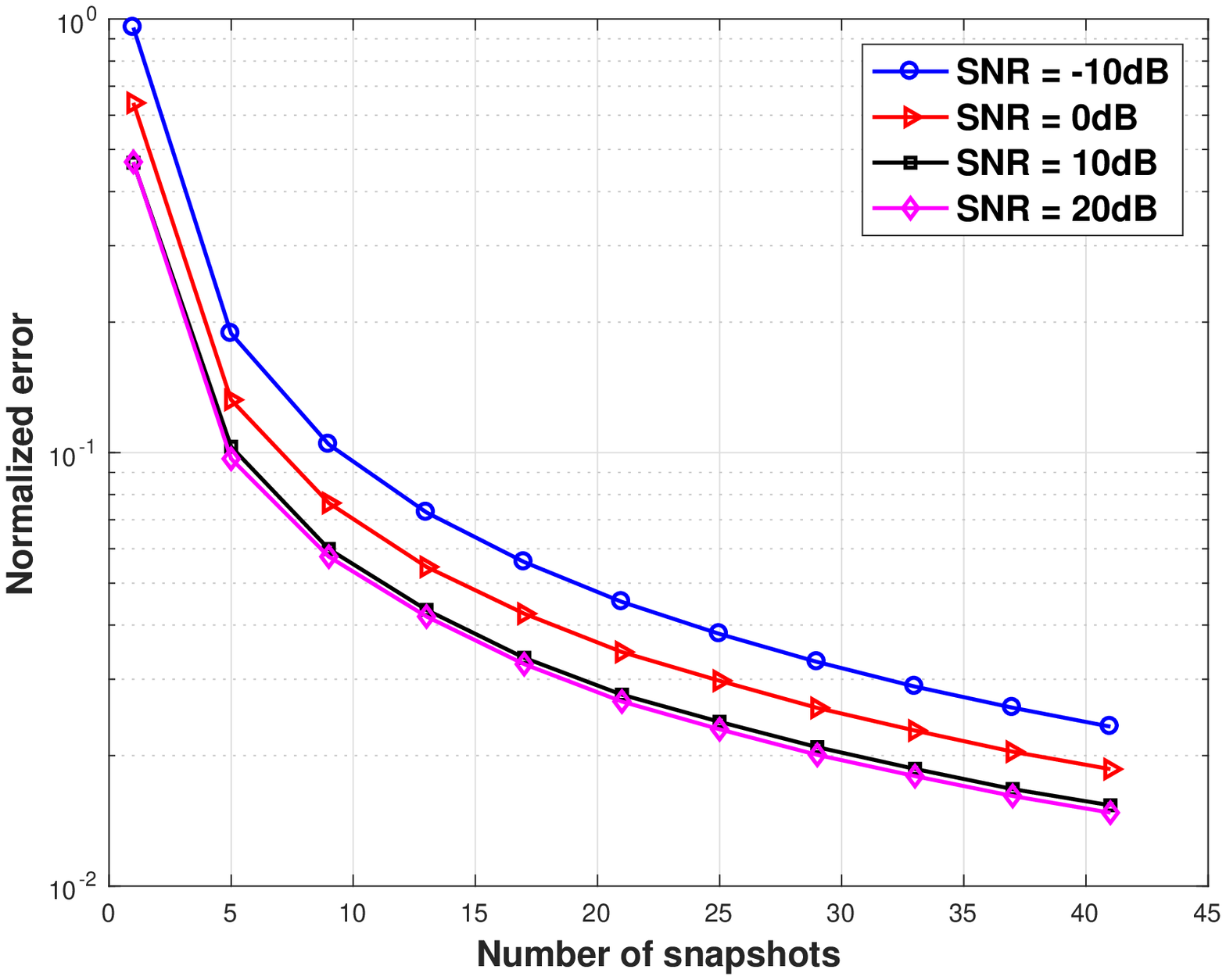}  &\includegraphics[width=0.45\textwidth]{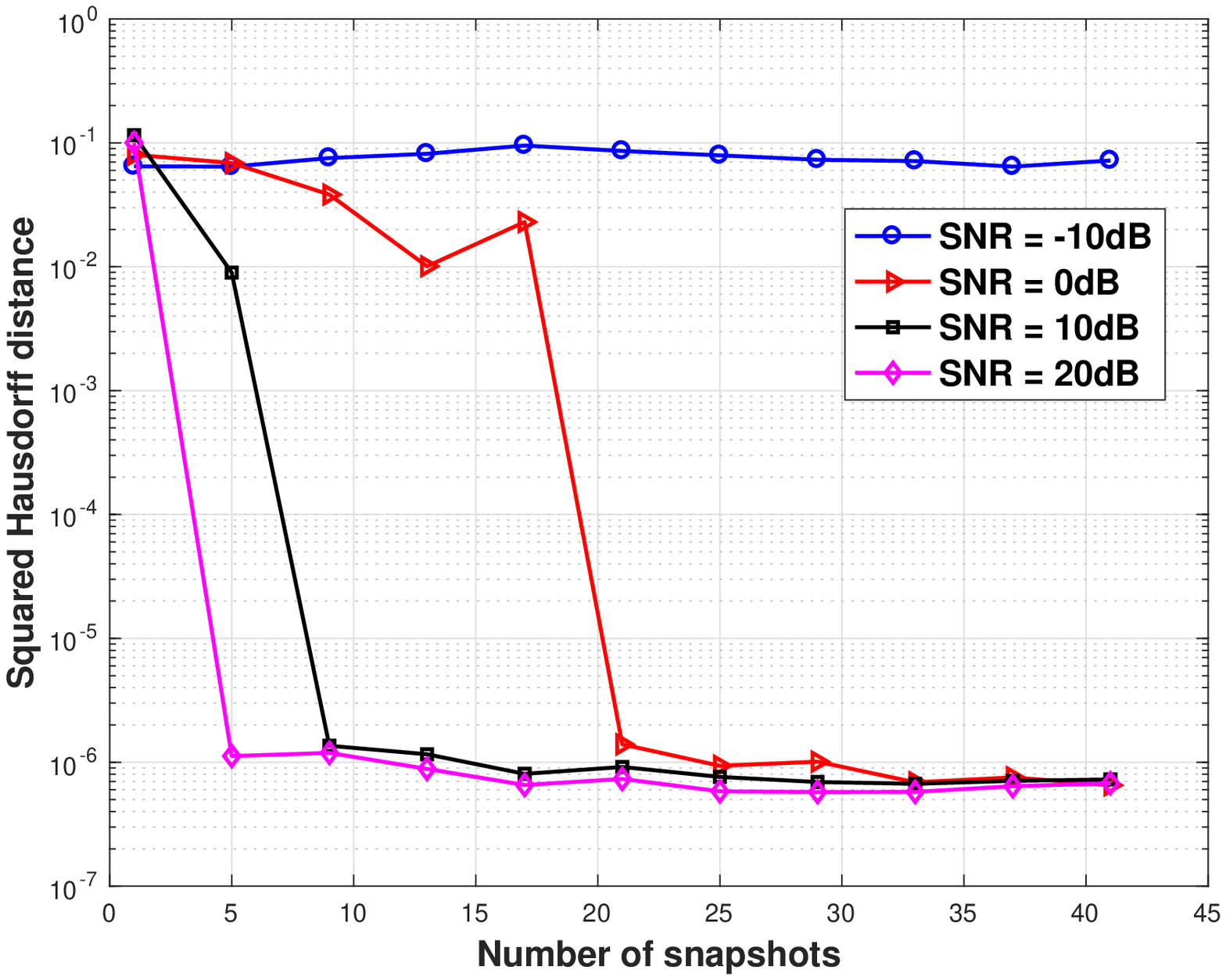}  \\
(a)   & (b)  
\end{tabular}
	\caption{Performance with respect to the number of snapshots at different SNRs using 1-bit measurements: (a) signal reconstruction error; (b) frequency estimation error measured in Hausdorff distance.}\label{onebit_MMV_SNR_Normalized}
\end{figure*} 

\subsection{Multiple Vector Case}

We evaluate the performance of the AST algorithm \eqref{atomic_denoising_MMV} in the multiple vector case. We follow the same setup as Fig.~\ref{1bit_localization}, where $n=64$, the set of frequencies $\boldsymbol{f}=\left\lbrace 0.3,0.325,0.8 \right\rbrace$, and the number of measurements for each snapshot is $m=50$. The coefficients of each snapshot in $\bX$ is generated independently using the standard complex Gaussian distribution. The SNR {\em per snapshot} is defined as $\mathsf{SNR} = \|\bX\|_{F}^{2}/(T\sigma^{2} )$, where $T$ is the number of snapshots. We set the regularization parameter $\tau_T = \sqrt{n\log n/(10\cdot m T)}$ in the experiment. The normalized reconstruction error is defined as $\sin^2(\angle\bX;\hat{\bX})$, where $\hat{\bX}$ is the recovered signal containing multiple snapshots, and $\angle$ denotes the angle between the subspace spanned by $\bX$ and $\hat{\bX}$. Once $\hat{\bX}$ is obtained, we estimate the frequencies  by using the MATLAB function \texttt{rootmusic} by assuming the correct model order, that is $K=3$. The accuracy of frequency estimation is evaluated by examining the Hausdorff distance between the recovered frequencies $\hat{\boldsymbol{f}}$ and the ground truth $\boldsymbol{f}$ as 
\begin{equation*}
	d_{H} ( {\boldsymbol{f}},\hat{\boldsymbol{f}}  ) = \max\left\{ \sup_{ f\in  \boldsymbol{f}} \inf_{\hat{f}\in \hat{\boldsymbol{f}}} \|f-\hat{f}\|_{2}, \sup_{\hat{f}\in \hat{\boldsymbol{f}}}\inf_{f\in \boldsymbol{f}} \|f-\hat{f}\|_{2}  \right\}.  
\end{equation*}

Fig.~\ref{onebit_MMV_SNR_Normalized} shows the recovery performance with respect to the number of snapshots at different SNRs, averaged over $50$ Monte Carlo simulations, where (a) depicts the normalized reconstruction error, and (b) depicts the squared Hausdorff distance. At a fixed SNR, it can be seen that both the normalized reconstruction error and frequency estimation error reduce, highlighting the benefit of having multiple snapshots. In particular, having multiple snapshots allows better frequency recovery once the number of snapshots is large enough. Moreover, performance improves as we increase the SNR.

\section{Concluding Remarks}\label{sec_conclusion}

In this paper, we examined the effect of (heavy) quantization in spectral compressed sensing that is useful for understanding wideband spectral signal acquisition and processing. Our contributions are two-fold. We first derived the Cram\'er-Rao bound for parameter estimation with multiple complex sinusoids using quantized compressed linear measurements. This bound is instrumental in describing the trade-offs between bit depth and sample complexity at different SNR regimes. Such an estimation-theoretical perspective is independent of the algorithm and hasn't been exploited in the previous literature. Secondly, we developed algorithms for spectral-sparse signal recovery using quantized measurements via atomic norm minimization, which do not require knowledge of the quantizer in recovery. Under a mild separation condition, we establish that we can accurately recover a spectrally-sparse signal from the signs of $O(K \log n)$ random linear measurements. The proposed algorithm also can be extended to handle multiple signal snapshots. This generalizes the literature on one-bit compressed sensing to the important class of spectrally sparse signals using atomic norms, and we carefully examined the performance of the proposed algorithms via numerical experiments. 

An alternative convex relaxation for spectrally-sparse signal recovery is based on Hankel matrix enhancement and nuclear norm minimization \cite{chen2013robust,cai2016robust}. In the single vector case, instead of imposing the atomic norm regularizer as in \eqref{1bit_atomic}, one may consider
\begin{equation}\label{1bit_hankel}
\hat{\bx} = \argmin_{\bx\in\mathbb{C}^n} \frac{1}{2}\|\bx - \bs \|_2^2 + \tau_H \|\mathcal{H}(\bx)\|_*.
\end{equation}
Here, $\mathcal{H}(\bx)$ denotes a Hankel matrix given as
\begin{equation*}
\mathcal{H}(\bx) =\begin{bmatrix}
x_1 &  x_2  & &\\
x_2 &  & \iddots& \\
\vdots & \iddots& & \\
x_{n_1} & x_{n_1+1}& \cdots & x_{n}
\end{bmatrix},
\end{equation*}
where $n_1$ is set as $\lfloor n/2 \rfloor$ to make the matrix $\mathcal{H}(\bx) $ as square as possible, $\|\cdot\|_*$ is the nuclear norm, and $\tau_H$ is a regularization parameter. Our preliminary numerical simulations suggest this method is also effective for promoting spectral sparsity, but a detailed study is beyond the scope of the current paper. We leave the thorough analysis of \eqref{1bit_hankel} to future work.

Since the Cram\'er-Rao bounds assume perfect knowledge of the quantizers, they may not be indicative to benchmark the performance of the atomic norm minimization algorithms as proposed in this paper, since these algorithms do not make use of such knowledge. In the future, it might be interesting to develop estimation-theoretical bounds that only assume partial or little knowledge about the quantizer.

\appendices

\section{Proof of Theorem~\ref{theorem_main}}\label{proof_theorem_smv}

An alternative way to represent the atomic decomposition is to write it as an integration of certain point measure \cite{tang2015near}. Define the representing measure of $\bx^{\star}$ as
\begin{equation*}
{\mu}(f) = \sum_{k=1}^K c_k \delta(f-f_k),
\end{equation*}
where $\delta(\cdot)$ is the delta function. Then we can rewrite $\bx^{\star}$ as
\begin{equation}
\bx^{\star} = \int_{0}^1 \bv(f)d \mu(f) = \sum_{k=1}^Kc_k \bv(f_k) .
\end{equation}
Correspondingly, denote $\hat{\mu}(f)$ as the representing measure for the solution $\hat{\bx}$ of \eqref{1bit_atomic}, which means $\hat{\bx} =\int_{0}^1 \bv(f)d\hat{\mu}(f) $.
 
Denote the reconstruction error as $\be = \lambda \bx^{\star} - \hat{\bx}$, and its representing measure is $\gamma = \lambda\mu - \hat{\mu}$. With these definitions, applying \cite[Lemma 1]{tang2015near}, we can bound the error as \cite{tang2015near}
 \begin{equation}\label{errorbound}
 	\| \be \|_{2}^{2} \leq \|\be \|_{\cA}^* \left( \int_{F}|\gamma |\left(df \right) + I_{0}+I_{1}+I_{2}  \right), 
 \end{equation} 
where $I_{\ell} = \sum_{k=1}^K I_{\ell}^k$, for $\ell=0,1,2$, with $I_{0}^{k}= \left|\int_{N_k}\gamma\left(df \right) \right|$, $I_{1}^{k}= n \left|\int_{N_{k}}\left(f-f_{k} \right) \gamma\left(df \right) \right|$, $I_{2}^{k}= \frac{n^{2}}{2} \int_{N_{k}}\left(f-f_{k} \right)^{2} |\gamma|\left(df \right) $, where $N_{k} = \left\lbrace f\in \mathbb{T}:d\left(f,f_{k} \right)\leq 0.16/n  \right\rbrace $ as the neighborhoods around each frequency, and $F = \mathbb{T}\setminus\cap_{k=1}^{K}N_{k}$. 

To bound the first term in \eqref{errorbound}, let us denote the deviation
 \begin{equation}
 \bw = \bs -\mathbb{E}[\bs] = \bs - \lambda \bx^{\star},
 \end{equation}
where $\mathbb{E}[\bw]=0$. We have
\begin{align}
	\|\be \|_{\cA}^* &\leq   \|\bw\|_{\cA}^*+  \|\bs - \hat{\bx} \|_{\cA}^*  \nonumber \\
	&\leq\|\bw\|_{\cA}^* + \tau, \label{bound_dual_e}
\end{align}  
where the first line follows from the triangle inequality, and the second line follows from the optimality condition of the AST algorithm in \eqref{1bit_atomic} in the following lemma.
\begin{lemma}[Optimality conditions \cite{bhaskar2013atomic}] \label{lemma_opt_condition}$\hat{\bx}$ is the solution of \eqref{1bit_atomic} if and only if $\| \bs - \hat{\bx}\|_{\cA}^*\leq \tau$, and $\langle \bs - \hat{\bx},\hat{\bx}\rangle = \tau \|\hat{\bx}\|_{\cA}$.
\end{lemma}    
Therefore, if we set $\tau\geq \eta \|\bw\|_{\mathcal{A}}^{*}$, where $\eta\geq 1$ is some constant, then plugging this into \eqref{bound_dual_e} we can show that 
\begin{equation} \label{dual_atomic_norm_e_bound}
\|\be\|_{\cA}^* \leq (\eta^{-1}+1)\tau \leq 2\tau.
\end{equation}

The second term in \eqref{errorbound} can be bounded in exactly the same manner as in \cite{tang2015near}, as long as \eqref{dual_atomic_norm_e_bound} holds. In effect, \cite{tang2015near} proved the following bound, under the separation condition, with high probability we have
\begin{equation} \label{bound_second_term}
 \left( \int_{F}|\gamma |\left(df \right) + I_{0}+I_{1}+I_{2}  \right) \leq C \frac{K\tau}{n}.
\end{equation} 

The following lemma bounds $\|\bw\|_{\cA}^*$, whose proof is provided in Appendix~\ref{proof_lemma2}.
\begin{lemma}\label{dual_atomic_norm_w}
With probability at least $1-1/(\pi n \log n)$, we have 
	\begin{equation*}
		\|\bw\|_{\cA}^* \leq C\cdot \sqrt{\frac{n\log n}{m}},  
	\end{equation*}
where $C$ is some universal constant.
\end{lemma}    

Therefore, set $\tau =C \eta \sqrt{n\log n/m}$, and plug \eqref{dual_atomic_norm_e_bound} and \eqref{bound_second_term} into \eqref{errorbound}, we have
\begin{equation}
\|\be\|_2^2 \leq C'  \cdot \frac{K\tau^2}{n} \leq C'\frac{K\log n}{ m}.
\end{equation}
which is equivalent to
\begin{equation*}
\left\|\frac{\hat{\bx}}{\lambda} - \bx^{\star} \right\|_{2} \lesssim \frac{1}{\lambda}\sqrt{\frac{K\log n}{m}}.
\end{equation*}
The proof is complete.

\section{Proof of Lemma~\ref{dual_atomic_norm_w}} \label{proof_lemma2}

By definition, we can write $\|\bw\|_{\mathcal{A}}^{*} $ as 
\begin{align}
\|\bw\|_{\mathcal{A}}^{*} &= \sup_{f\in[0,1)} \left|\langle \bs-\lambda \bx^{\star}, \bv\left(f \right)  \rangle \right| \nonumber \\
&= \sup_{f\in[0,1)}  \left|\langle \bs,\bv\left( f\right)   \rangle - \mathbb{E}\left[\left\langle \bs,\bv\left( f\right)   \right\rangle \right]  \right| \nonumber \\
& = \sup_{f\in[0,1)} \left| g_{\bx^{\star}}(f)  - \mathbb{E}[g_{\bx^{\star}}(f) ] \right| \label{w_reformulation}
\end{align}
where 
$ g_{\bx^{\star}}(f) : = \left\langle \bs,\bv\left( f\right)   \right\rangle =  \frac{1}{m} \sum_{i=1}^{m}y_{i}\left\langle \ba_{i},\bv\left(f \right)  \right\rangle$. 	

To proceed, we use the following symmetrization bound, which is the complex-valued version of \cite[Lemma 5.1]{plan2013robust}.   
\begin{lemma}\label{Complex_symmetric}
	Let $\left\lbrace\epsilon_{i}\right\rbrace_{i=1}^{m} $ be a sequence of independent complex-valued random variables, where $\epsilon_i\sim \epsilon= e^{j2\pi \theta}$, where $\theta$ uniformly distributed between $[0,1)$. Then 
	\begin{align} \label{expectation}
	\mu :&= \mathbb{E}\left[ \sup_{f\in[0,1)} \left|g_{\bx^{\star}}\left(f \right)  - \mathbb{E}\left[g_{\bx^{\star}} (f  )  \right] \right|  \right] \nonumber \\
	&\quad\quad\quad \leq 2\mathbb{E}\left[\sup_{f\in[0,1)} \frac{1}{m}\left|\sum_{i=1}^{m}\epsilon_{i}y_{i}\left\langle\ba_{i},\bv (f  )  \right\rangle \right| \right].  
	\end{align} 
Furthermore, we have the deviation inequality
	\begin{align} \label{deviation}
	&\mathbb{P}\left\lbrace \sup_{f\in[0,1)}   \left|g_{\bx^{\star}}\left(f \right)  - \mathbb{E}\left[g_{\bx^{\star}}\left(f \right)\right] \right|  \geq 2\mu +t  \right\rbrace \nonumber \\ 
	&\quad\quad\quad\leq 4 \mathbb{P}\left\lbrace \sup_{f\in[0,1)} \frac{1}{m} \left|\sum_{i=1}^{m}\epsilon_{i}y_{i}\left\langle\ba_{i},\bv\left(f \right)  \right\rangle \right|>\frac{t}{2}  \right\rbrace.  
	\end{align}
\end{lemma}

Before applying Lemma~\ref{Complex_symmetric}, note that by symmetrization and rotational invariance, $\epsilon_i y_i \ba_i$ have the same i.i.d. distribution of $\sqrt{2}\ba_i$. Therefore, the following quantities are equivalent in distribution:
\begin{align*}
\sup_{f\in[0,1)} \frac{1}{m}\left |\sum_{i=1}^{m}\epsilon_{i}y_{i}\left\langle\ba_{i},\bv\left(f \right)  \right\rangle \right| 
&\sim \frac{\sqrt{2}}{m} \sup_{f\in[0,1)}  \left| \sum_{i=1}^m \left\langle\ba_i ,\bv\left(f \right)  \right\rangle \right|  \\
& \sim \sqrt{\frac{2}{m}} \underset{f\in[0,1)}{\mathrm{sup}} |\left\langle\bg,\bv\left(f \right)  \right\rangle|,    
\end{align*}  
where $\bg$ is a vector composed of i.i.d. $\mathcal{CN}(0,1)$.

Applying \eqref{deviation} in Lemma~\ref{Complex_symmetric} to \eqref{w_reformulation}, we have
\begin{equation}\label{Up_Dual_w}
\mathbb{P}\left(\|\bw\|_{\mathcal{A}}^{*} \geq 2\mu + t \right) \leq 4\mathbb{P}\left(\sqrt{\frac{2}{m}} \sup_{f\in[0,1)}|\left\langle\bg,\bv\left(f \right)  \right\rangle| \geq \frac{t}{2} \right).  
\end{equation}  

From \eqref{expectation} in Lemma~\ref{Complex_symmetric}, we have
\begin{align}\label{bound_for_mu}
\mu = \mathbb{E}\left[ \|\bw\|_{\mathcal{A}}^{*}\right] & \leq 2\sqrt{\frac{2}{m}} \mathbb{E}\left[ \sup_{f\in[0,1) } |\left\langle\bg,\bv\left(f \right)  \right\rangle| \right] \nonumber \\
& \leq C\sqrt{\frac{n\log n}{m}},
\end{align} 
where the second line follows from \cite[Appendix C,D]{bhaskar2013atomic} as
\begin{equation*} 
\mathbb{E}\left[\underset{f\in[0,1]}{\mathrm{sup}} |\left\langle\bg,\bv\left(f \right)  \right\rangle| \right]  \leq  C_1 \sqrt{n\log\left(n \right)  }. 
\end{equation*} 

Moreover, from \cite[Appendix C]{bhaskar2013atomic}, we have
\begin{equation*}
\sup_{f\in[0,1)}  \left|\left\langle\bg,\bv\left(f \right)  \right\rangle \right| \leq C_2 \cdot \sqrt{n\log n}
\end{equation*}
hold with probability at least $1-1/(\pi n \log n)$. Set $t=2C_2\sqrt{n\log n}$ and plug in the above two inequalities in \eqref{Up_Dual_w}, we have that
\begin{equation*}
\|\bw\|_{\mathcal{A}}^{*} \leq C\cdot \sqrt{\frac{n\log n}{m}}  
\end{equation*}
holds with probability at least $1-1/(\pi n \log n)$.

\bibliographystyle{IEEEtran}
\bibliography{1bitAtomic}

\end{document}